\documentclass[12pt,preprint]{aastex}

\newcommand{\inst}{\altaffilmark}

\shorttitle{High energy properties of PKS~1830-211}
\shortauthors{Shu Zhang et al.}
\begin{document}
   \title{High energy properties of PKS~1830-211}
    \author {Shu Zhang \inst{1}, Yu-peng Chen \inst{1}, Werner Collmar \inst{2}, Luigi Foschini \inst{3}, Ti-Pei Li \inst{1,4}, \\
Diego F. Torres \inst{5}, Jian-Min Wang \inst{1,6}}

  \altaffiltext{1} {Key Laboratory for Particle Astrophysics, Institute of High Energy Physics, 19B YuQuan Road, Beijing 100049, China}
  \altaffiltext{2} {Max-Planck-Institut f\"ur Extraterrestrische Physik, P.O. Box 1603, 85740 Garching, Germany}
   \altaffiltext{3} {INAF/IASF-Bologna, via Gobetti 101, 40129, Bologna, Italy }
   \altaffiltext{4} {Center for Astrophysics,Tsinghua University, Beijing 100084, China}
\altaffiltext{5} {ICREA \& Institut de Ci\`encies de l'Espai (IEEC-CSIC), Campus UAB,
Facultat de Ci\`encies, Torre C5-parell, 2a planta, 08193 Barcelona,
Spain }
\altaffiltext{6} {Theoretical Physics Center for Science Facilities (TPCSF), CAS.}             

\begin{abstract} 
We report on an analysis of X- and $\gamma$-ray observations of PKS~$1830-211$, based on the long-term campaigns carried out by \emph{INTEGRAL} and COMPTEL. The \emph{INTEGRAL} data currently available present a $33\sigma$ significance detection in the $20-100$~keV band, while the COMPTEL $6$-years data provide a $5.2\sigma$ significance detection in the $1-3$~MeV energy band. At hard X-rays, \emph{INTEGRAL} and supplementary \emph{SWIFT} observations show flux variability on  timescales of  months. At $\gamma$-rays, the source shows persistent emission over years. The hard X-ray spectrum is well represented by a power-law model, with $\Gamma \sim 1.3$ in the $20-250$~keV band. This photon index is well consistent with the previous report of $\Gamma \sim 1.3$ obtained at $E > 3.5$~keV from the best fit of \emph{XMM-Newton} data with a broken power law model. The joint \emph{XMM-Newton}/\emph{INTEGRAL} spectrum presented here is then fit with a broken power-law model and the parameters are refined compared to the previous. The results show the photon index changes from $\sim 1.0$ to $\sim 1.3$ at a break energy $\sim 4$~keV. At MeV energies, the spectrum softens to $\Gamma \sim 2.2$. These results, together with the EGRET measurement at $E \ge 100$~MeV, constitute a broad-band spectrum containing the peak of the power output at MeV energies, similar to most high-luminosity $\gamma$-ray blazars. The measured spectral characterstics are then discussed in the framework of the gravitational lens effects.
\end{abstract}


\section{Introduction}

Since the first gravitational lens candidate was detected in 1979 (Walsh et al. 1979), the total number of such system has  grown since (e.g. Schneider et al. 1992). Among them is the high redshift blazar PKS~$1830-211$ ($z=2.507$), gravitationally lensed by an intervening galaxy at $z=0.89$. The discovery of such gravitational system traces back to radio observations (Pramesh Rao \& Subrahmanyan 1988). The radio map was composed of two compact components separated by $1''$, supposed to be the split images from the central region of the source and an extended structure which is most probably from the jet and regarded as an unusually strong Einstein ring (see Jauncey et al. 1991; Sunita Nair et al. 1993). Steppe et al. (1993) have shown that PKS~$1830-211$ is radio variable on timescales of months.  

X-ray observations (\emph{Chandra} and \emph{INTEGRAL}) revealed a quite hard spectrum, modeled with a power-law with a photon index $1.09\pm 0.05$ over the energy band $0.5-80$~keV (De Rosa et al. 2005). The spectral flattening at low energies has been often modeled with absorption in excess to the Galactic column ($N_{\rm H}=2.05\times 10^{21}$~cm$^{-2}$, Kalberla et al. 2005). The column density has been measured at soft X-rays by \emph{ROSAT} (Mathur \& Nair 1997) and \emph{ASCA} (Oshima et al. 2001) as well as by \emph{Chandra} (De Rosa et al. 2005), which suggests a column density of either $\sim 10^{22}$~cm$^{-2}$ at the lensing galaxy ($z=0.86$) or  $\sim 10^{23}$~cm$^{-2}$ intrinsic to the source. \emph{XMM-Newton} observations gave similar results, but the best fit is obtained with a broken power-law model, with the photon index changing from $\sim 1.0$ to $\sim 1.3$ at energies around $3.5$~keV (Foschini et al. 2006).

PKS~$1830-211$ was included in the first \emph{INTEGRAL} catalogue by Beckmann et al. (2006), with a photon index $1.96^{+0.27}_{-0.24}$ in the $20-100$~keV energy band and in the \emph{INTEGRAL} extragalactic survey by Bassani et al. (2006), with a $20-100$~keV flux $\sim 3$~mCrab level averaged over the first $2.5$ years of the \emph{INTEGRAL} observations.

At MeV energies PKS~$1830-211$ was firstly reported by Collmar (2006). The COMPTEL first 4-year observations (1991-1995) revealed a detection of $4.5\sigma$ in the $1-3$~MeV band. Contemporaneously, PKS~$1830-211$ was detected by EGRET at $\ge 100$~MeV with $7.8\sigma$ significance and a photon index $2.59 \pm 0.13$ (Hartman et al. 1999).

Since the amount of \emph{INTEGRAL} public data has been significantly increased after the last report, we decided to carry out a detailed analysis of PKS~$1830-211$ with all available \emph{INTEGRAL} data. We also reanalyzed the \emph{XMM-Newton} data with a newer software version, which allow us to extend the analysis down to $0.2$~keV. The data from COMPTEL and EGRET are also reanalyzed and added in order to obtain the best broad-band high-energy spectrum available to date. 

Long-term observations by \emph{SWIFT} at hard X-rays and by COMPTEL at MeV energies are dug for investigating the spectral and flux variability.  The results are finally discussed in the context of a gravitational lensing system, as is appropriate to PKS~$1830-211$.

In the following we assume $H_0=73.4$~km~s$^{-1}$~Mpc$^{-1}$ and $q_0=0$, measured from the latest \emph{WMAP} data (Spergel et al. 2007).

\section{Observations and data analysis}
\subsection{\emph{INTEGRAL}}

\emph{INTEGRAL} is an ESA scientific mission dedicated to high-resolution spectroscopy ($E/\Delta E \simeq 500$; SPI see Vedrenne et al. 2003) and imaging (angular resolution: $12'$ FWHM, point source location accuracy: $\simeq 1'-3'$; IBIS, see Ubertini et al. 2003) of celestial $\gamma$-ray sources in the energy range $15$~keV to $10$~MeV, with simultaneous monitoring at X-rays ($3-35$~keV, angular resolution: $3'$; JEM-X, see see Lund et al. 2003) and optical wavelength (Johnson V-filter, $550$~nm; OMC, see see Mas-Hesse et al. 2003). All the instruments onboard  \emph{INTEGRAL}, except OMC, work with coded masks. The observational data from the detector IBIS/ISGRI ($20-250$~keV, Lebrun et al. 2003) have been considered in our analysis of PKS~$1830-211$, because of highest quality.

The available \emph{INTEGRAL} observations when PKS~$1830-211$ fell into the Fully-Coded Field of View (FCFoV) of ISGRI (up to April 29, 2006; see Table \ref{tab:isgri-obs}) comprise about $1095$ science windows (scw), for a total exposure of $2500$~ks, i.e., about $550$~ks of new data analyzed here for the first time. Most of these observations were carried out in a $5\times 5$ dithering mode. The analysis were performed by using the \emph{INTEGRAL} \texttt{Offline Scientific Analysis (OSA) version 7.0}, whose algorithms for IBIS are described in Goldwurm et al. (2003). All the sources within the FOV which are brighter or comparable to PKS~$1830-211$ were taken into account in extracting the source spectrum and light curve. An additional $3$\% systematic error was added to the spectra because of calibration uncertainties. The spectra were fitted with \texttt{XSPEC v 12.3.1} and the model parameters were estimated at $90$\% confidence level.

\begin{figure}[ptbptbptb]
\centering
 \includegraphics[angle=0, scale=0.7]{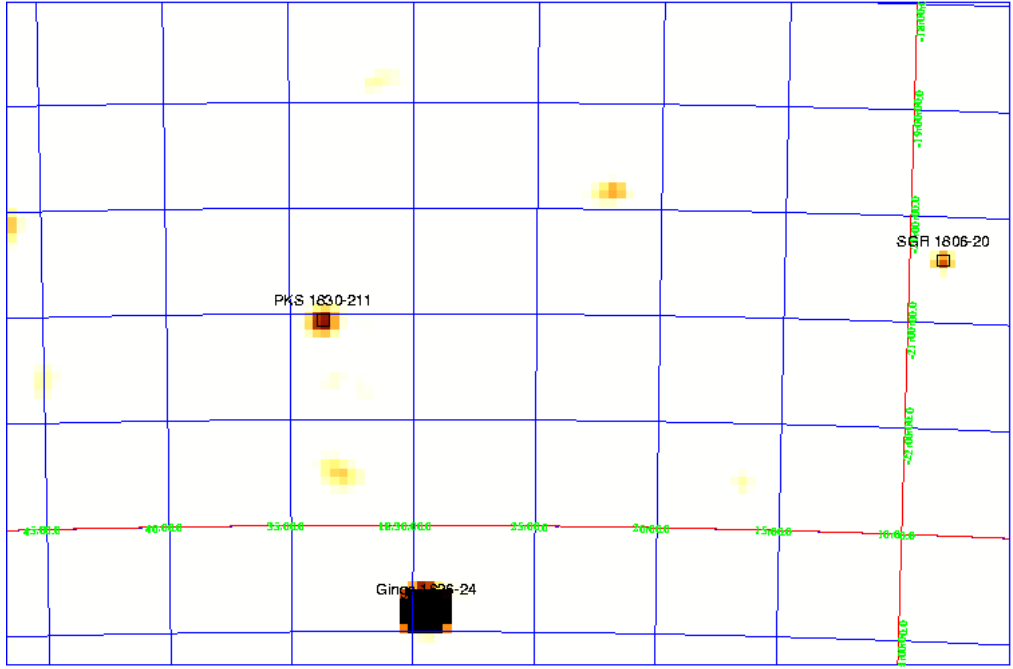}
      \caption{The ISGRI significance map of PKS~$1830-211$ region in the $20-100$~keV band, obtained by combining the observations of $2003 - 2006$.}
\label{skymap}
\end{figure}

The sum of all observations provide a detection with IBIS/ISGRI at $\sim 33\sigma$ level in the $20-100$~keV energy band (Fig.\ref{skymap}). The average $20-100$~keV flux is $0.75\pm 0.03$ counts/s, corresponding to $\sim 3$~mCrab, consistent with the results obtained by Bassani et al. (2006) over the first 2.5-year data.

The spectrum of the entire ISGRI data is well fit ($\tilde{\chi}^2 = 0.59$ with $8$ degrees of freedom) by a power-law model with $\Gamma = 1.29^{+0.16}_{-0.15}$ in the $20-250$~keV energy band.  We would like to note that the contribution from 150-250 keV is rather small and therefore does not affect the overall spectral fitting.

\subsection{\emph{XMM-Newton}}
The \emph{XMM-Newton} data available for PKS~$1830-211$ are three observations with the ObsID $0204580201$, $0204580301$ and $0204580401$, carried out in March $10$, $24$, and April $05$, 2004, respectively (Table~ \ref{tab:xmm-obs}). Data of EPIC-PN (Str\"uder et al. 2001) and EPIC-MOS (Turner et al. 2001) have been processed, screened and analyzed by using the same procedures described in Foschini et al. (2006), but with \texttt{SAS v 7.1.0} with the latest calibration files as of July $16$, $2007$. This software version allows us to probe the lowest energy part of the spectrum (down to $0.2$~keV). 

The results of the fits on the individual and averaged ObsID are summarized in Table~ \ref{tab:fitxmm}. The three observations show no significant flux and spectral variability; therefore, we integrated, with the \texttt{FTOOL addspec}, all the MOS1, MOS2 and PN data, respectively, to obtain an average spectrum for the joint fit with ISGRI data. Although, the broken power-law and the log-parabola models can be fit with similar results in the $\chi^2$ test, the broken power-law is slightly better and we consider this model as the best fit. The use of a ionized absorber intrinsic to the source, to model the low-energy photon deficit, gave worst results ($\tilde{\chi}^2=1.14$ for $1397$ dof in the average spectrum).  A change in ${\chi}^2$ is about 130 with respect to the cold absorption shown in Table~ \ref{tab:spec-fit}.

\subsection{\emph{SWIFT}}

\emph{SWIFT} is a $\gamma$-ray burst explorer and was launched on November 20, 2004. It carries three co-aligned detectors (Gehrels et al. 2004), namely the Burst Alert Telescope (BAT, Barthelmy et al. 2005), the X-Ray Telescope (XRT, Burrows, et al. 2005) and the Ultraviolet/Optical  Telescope (UVOT, Roming et al. 2005). 

BAT has rather large field of view of $1.4$~sr in partially-coded mode and works in the $15-150$~keV energy band. This makes it possible for a source to be daily monitored at the hard X-rays. The data products are therefore the source lightcurves and are publicly available\footnote{See the \emph{SWIFT}/BAT transient monitor results provided by the \emph{SWIFT} Team at \texttt{http://swift.gsfc.nasa.gov/docs/swift/results/transients/}}. The BAT lightcurve traces PKS~$1830-211$ back to February 12, 2005, in the $15-50$~keV energy band (see Fig.\ref{swift_lc}). This lightcurve has a weighted-average flux value of $4.56\pm 0.48 \times 10^{-3}$~counts~cm$^{-2}$~s$^{-1}$, corresponding to $2$~mCrab at $\sim 9.5\sigma$ over a time period of roughly $2.5$ years.

\subsection{COMPTEL}
The imaging Compton Telescope COMPTEL (1991-2000) onboard the \emph{Compton Gamma-Ray Observatory} (CGRO) was sensitive to $\gamma$-rays in the $0.75-30$~MeV energy range with an energy resolution of $\sim 10$\%. It had a large field of view and was able to detect $\gamma$-ray sources with an accuracy of the order of $1^{\circ}-2^{\circ}$ (e.g., see Sch\"onfelder et al. 1993). 

The standard imaging method of maximum likelihood was applied for COMPTEL data analysis. The detection significance can be estimated from the quantity $-2 \ln \lambda$, where $\lambda$ is the ratio of the likelihood $L$ for the background and the source plus background. For a known source, $-2 \ln \lambda$ has the $\chi^2$ distribution with $1$ free parameter in addition to the null hypothesis ($\chi_{1}^{2}$, de Boer et al. 1992). The point spread function of the instrument is applied by assuming an E$^{-2}$ power law shape for the input spectrum. The background is derived, with the first order of approximation, by a filter technique in data space (Bloemen et al. 1994). 

\begin{figure}[ptbptbptb]
\centering
 \includegraphics[angle=0, scale=0.7]{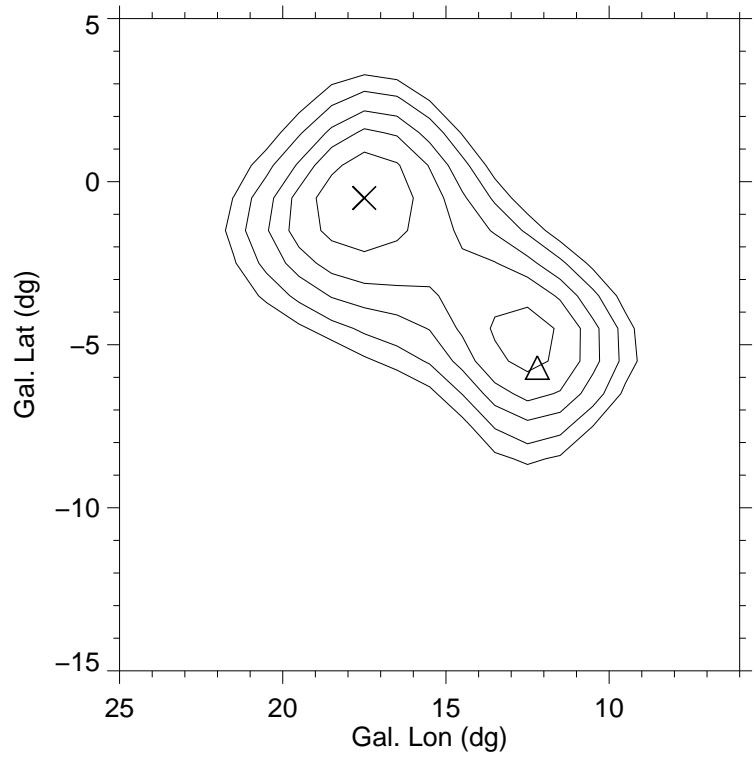}
      \caption{The $1-3$~MeV map from COMPTEL observations of phases $1-6$. The contours start at detection significance level of $3\sigma$, with the step of $1\sigma$. The symbols of the triangle represent the location of PKS~$1830-211$, and the cross the location of the so-called MeV source $l=18$.}
         \label{mevmap}
\end{figure}

PKS~$1830-211$ was marginally detected in the $1-3$~MeV band at $4.5\sigma$ level, using the first 4-years data (Collmar 2006). Here we take the complete 6-years COMPTEL data (see Table \ref{tab:mev-obs}), until the second reboost of the satellite in 1997, after which the background changed a lot making it difficult for further research, to investigate again the MeV emission. These data are subdivided into the so-called \emph{CGRO} phases, with each period covering typically one year of observations. The source is again detected mainly in $1-3$~MeV band, but the detection significance is improved to $5.2\sigma$ ($-2 \ln \lambda \sim 27$) by using two-year more data. Fig.~\ref{mevmap} shows the sky map in $1-3$~MeV band. The source fluxes are given in Table~\ref{tab:mev-flux}, in $4$ energy bands ($0.75-1$, $1-3$, $3-10$ and $10-30$ MeV, respectively).

\section{Time variability}

\begin{figure}[ptbptbptb]
     \centering
     \includegraphics[angle=0, scale=0.7]{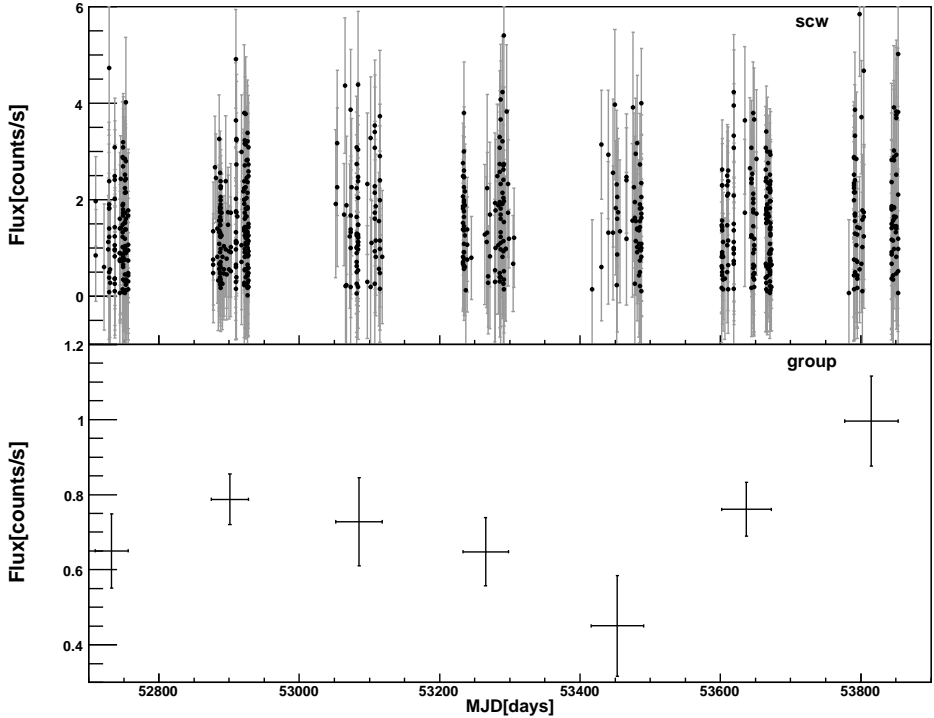}
     \caption{ISGRI light curves in the $20-100$~keV band, on basis of scw (top) and observational group (bottom), for the time period  between $2003$ and $2006$.}
		     \label{isgri_lc}
  \end{figure}

\begin{figure}[ptbptbptb]
\centering
\includegraphics[angle=0, scale=0.7]{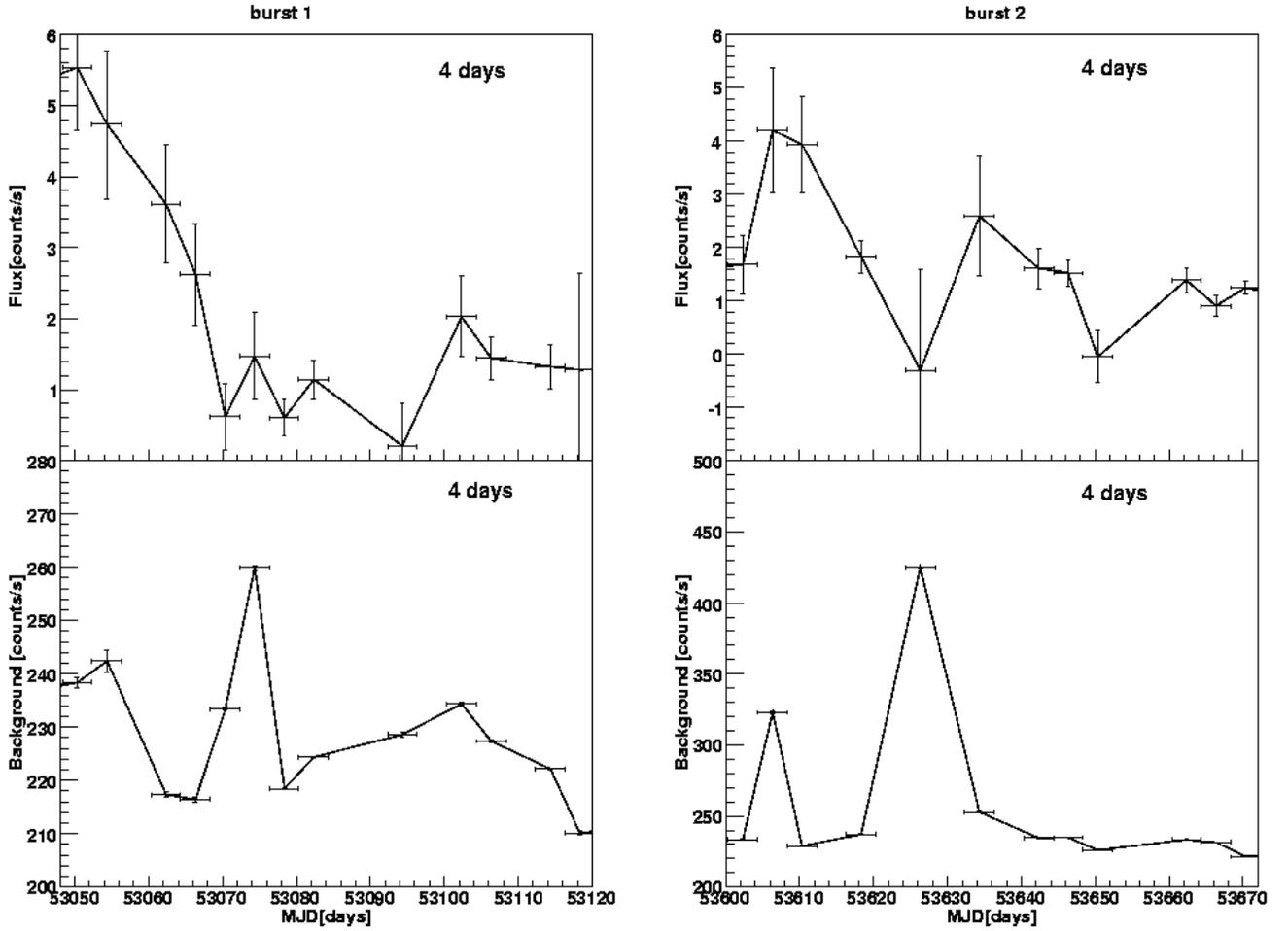}
\caption{Two high-variability episodes detected by ISGRI in the $20-100$~keV energy band, with each bin representing $4$ days. The lower panels show the corresponding background light curves.}
\label{bursts}
\end{figure}

The ISGRI light curve on a scw basis shows no clear flux variability (see Fig.~\ref{isgri_lc}). A fit to this lightcurve with a constant results in a $\tilde{\chi}^2 \sim 0.72$. To improve the statistics, data from each observational group (totally there are 7 groups, separated by long observational gaps) are combined to produce alternative lightcurves, as also shown in Fig.~\ref{isgri_lc}. The flux tends to drop smoothly during the first $2.5$ years since 2003 and then to raise in the following years. A search for the  flux variability on shorter time scales (4 days bins) resulted in two interesting episodes (see Fig.~\ref{bursts}). In the first episode, the flux drops by a factor of about 6 on a time scale of about 20 days, while in the second episode (this one has low significance), the flux changed by a factor 2 on time scale of 8 days.  Such excess in flux, although weak, is indicated as well in the energy bands of 20-40 keV and 40-100 keV. However, these episode events might be regarded only as hints for flux variability and the statistics is not sufficient to investigate in details these episodes.

\begin{figure}[ptbptbptb]
   \centering
 \includegraphics[angle=0, scale=0.7]{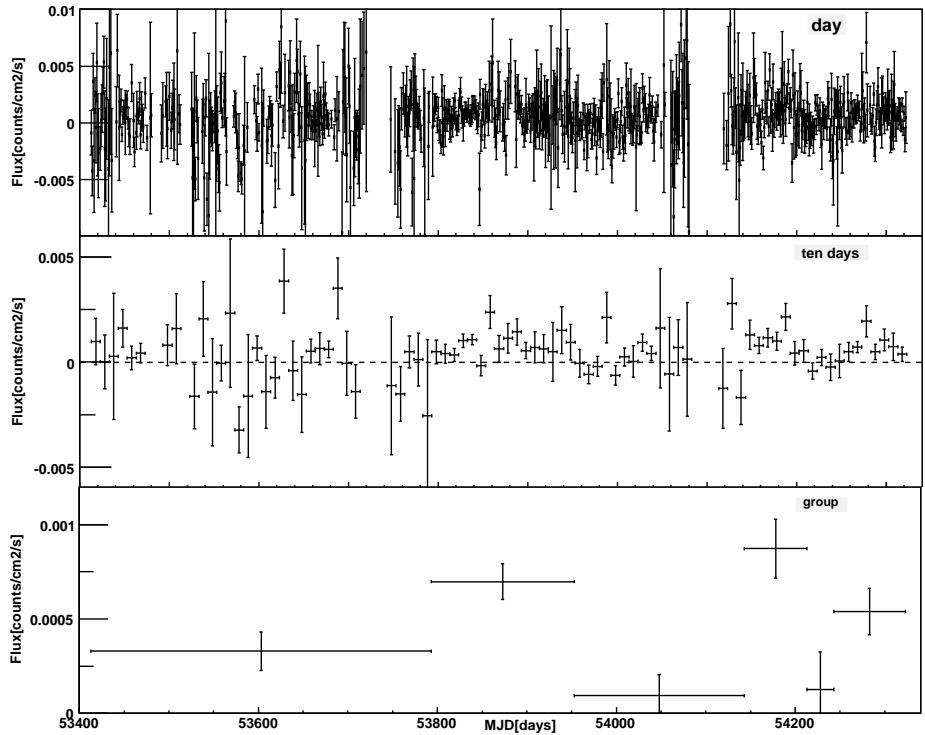}
      \caption{\emph{SWIFT} light curves ($15-50$~keV) with each bin representing a time scale of one day (top), ten days (middle) and an observational group (bottom). The observational groups are defined to have persistent emission excess as showing up in the lightcurve of the ten-day basis (the middle panel).}
          \label{swift_lc}
\end{figure}

\emph{SWIFT}/BAT provides a daily lightcurve in the $15-50$~keV energy band since February 2005. However, large error bars, mainly due to systematics, prevent us from inferring trends in flux evolution. Therefore, the data are combined in $10$-days bins and the resulted lightcurve shows three time zones with persistent flux excess (see Fig. \ref{swift_lc}). Accordingly, the observations are divided again into 6 parts, over which the weighted average in flux are shown in Fig.\ref{swift_lc}. This lightcurve suggests that PKS~$1830-211$ is rather variable at the hard X-rays on the time scale of months.

\begin{figure}[ptbptbptb]
     \centering
     \includegraphics[angle=0, scale=0.9]{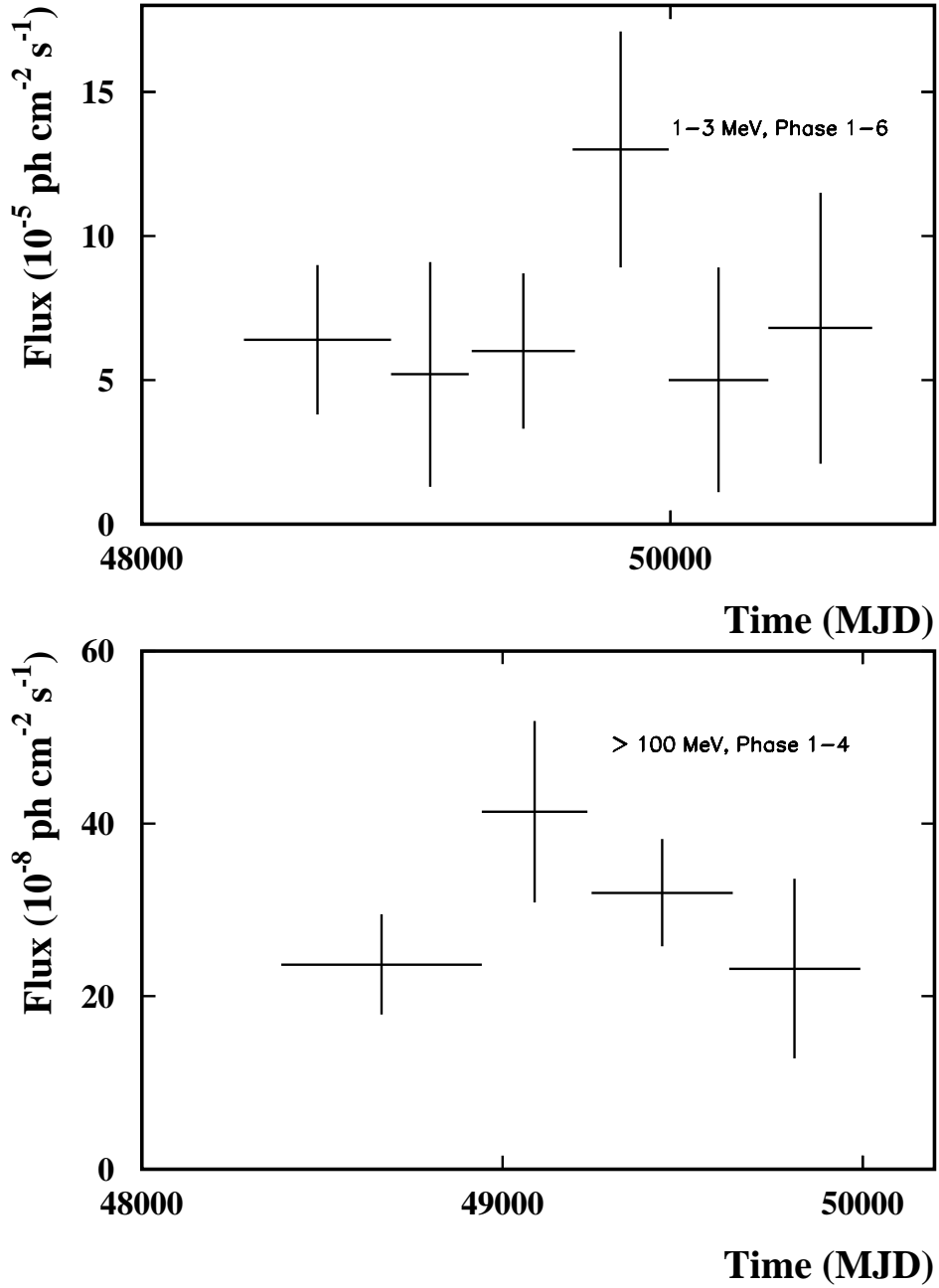}
     \caption{COMPTEL (upper panel) and EGRET (lower panel) lightcurves with each bin averaged over one \emph{CGRO} phase. The error bars are $1\sigma$. }
     \label{mevlc}
  \end{figure}

The lightcurves from COMPTEL ($1-3$~MeV, time period $1991-1997$) and EGRET ($\ge 100$~MeV, time period $1991-1995$) are shown in Fig. \ref{mevlc}, with each bin presenting the average of one \emph{CGRO} satellite observing phase. These lightcurves indicate that PKS~$1830-211$ is likely to have persistent emission over years at $\gamma-$rays.

\section{Broad-band energy spectrum}

\begin{figure}[ptbptbptb]
  \includegraphics[angle=0, scale=0.7] {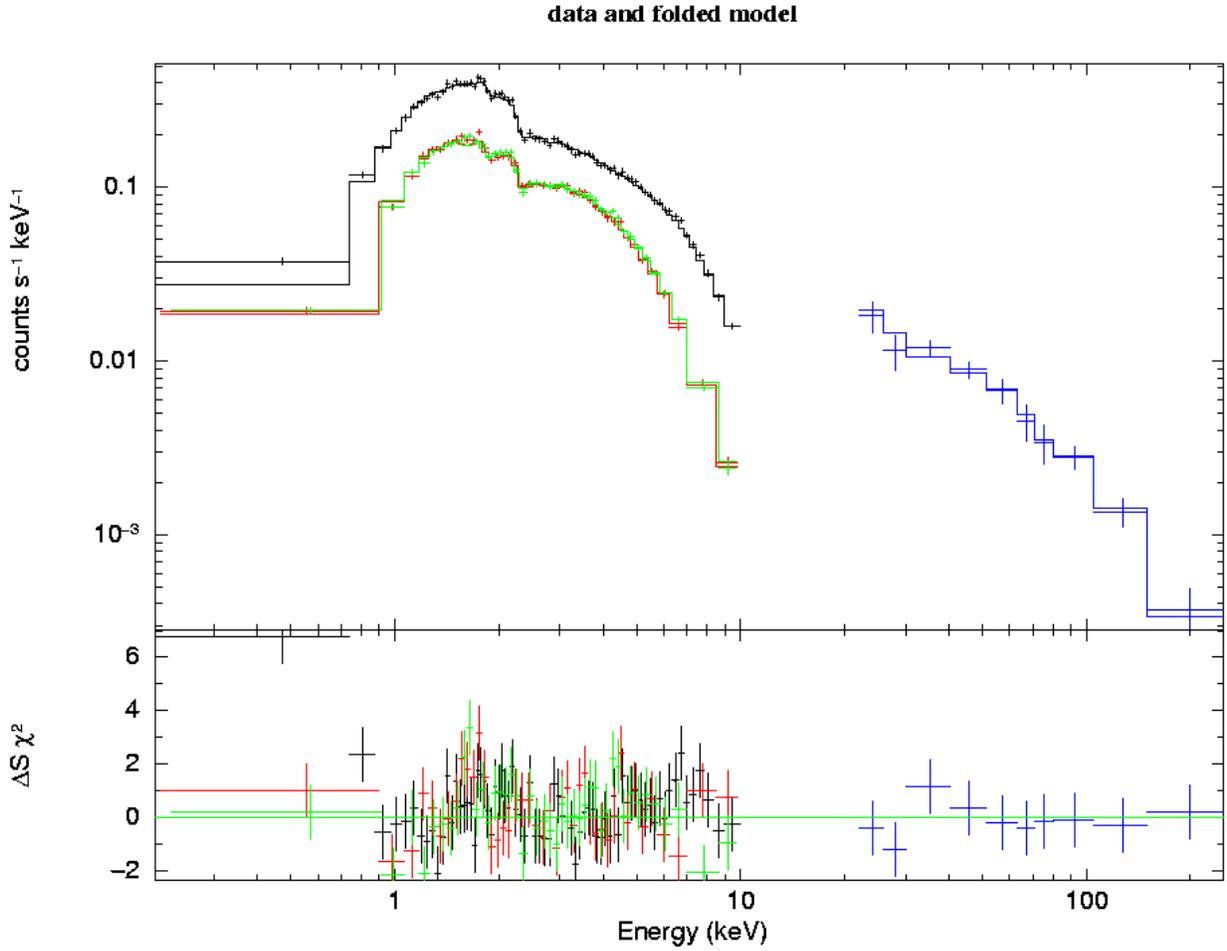}
     \caption{Spectral fit with a  broken power law model for  the combined  \emph{XMM-Newton} data (at energy below $10$~keV), and the ISGRI data (at energy above $20$~keV). For better visibility the \emph{XMM-Newton} spectra have been rebinned in the plot.}
         \label{spe}
   \end{figure}

We performed a joint fit of the \emph{XMM-Newton}/\emph{INTEGRAL} data. The results are shown in Table~ \ref{tab:spec-fit} and Fig.~\ref{spe}. The source luminosity in the $0.2-250$~keV energy band is calculated as $3.5 \times  10^{48}$~erg~s$^{-1}$. We note that the fit by a simple power-law model ($\Gamma \sim 1.12$) results in a reduced $\chi^2$ $\sim 1.13$ under $1408$ dofs. Therefore, the broken power-law model results to be again the best fit model, with an improvement with respect to the single power-law model $>99.99$\% as calculated with the \emph{F}-test.

At MeV energies the spectrum can be well represented ($\tilde{\chi}^2 = 0.4$ for $2$ dof) by a single power-law model with $\Gamma = 2.23^{+0.36}_{-0.27}$ measured from COMPTEL data combined from \emph{CGRO} phases 1-4. This spectrum is then used in the followings to compare with the simultaneous EGRET one in the $\ge 100$~MeV energy band (Hartman et al. 1999).  

\begin{figure}[ptbptbptb]
  \includegraphics[angle=0, scale=0.9] {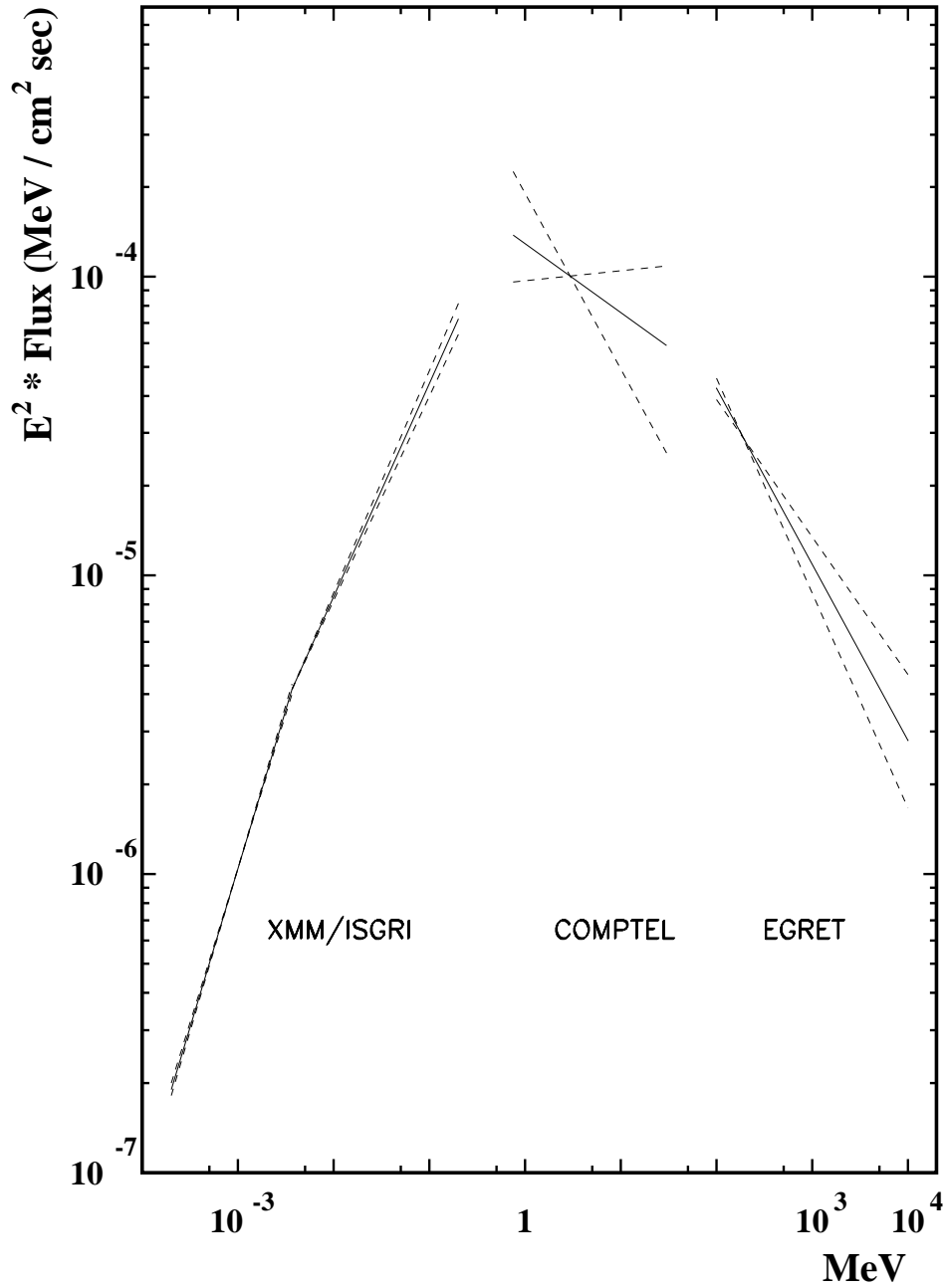}
     \caption{The broad energy spectrum of PKS~$1830-211$. The solid lines are the broken power law shape  as obtained by \emph{XMM-Newton}/\emph{INTEGRAL} at X-rays, power law shapes by COMPTEL at MeV energies, and by EGRET at $\ge 100$~MeV (Hartman et al., 1999). The dashed lines are the plots of the $1\sigma$ error in spectral shape.}
\label{sed}
\end{figure}

Fig.~\ref{sed} shows the high-energy broad-band spectrum for the results derived in this paper, including the data from \emph{XMM-Newton} and \emph{INTEGRAL} to cover the soft/hard X-rays ($0.2-250$~keV) energy band; from COMPTEL for the $\gamma$-rays ($0.75-30$~MeV); and from EGRET for the $\gamma-$rays in the $\ge 100$~MeV energy band (Hartman et al. 1999). We considered the EGRET and COMPTEL spectra obtained simultaneously in phases $1-4$, because they were  co-aligned in pointing onboard \emph{CGRO}.

In such a broad-band view, the power output of PKS~$1830-211$ shows a bump located at MeV energies, as expected in the common view of a high-luminosity blazar, where the high-energy part of the spectral energy distribution (SED) is due to inverse-Compton emission from the relativistic electrons in a jet scattering of seed photons coming from a source external to the jet (broad-line region, accretion disk,...; see Fossati et al. 1998, Ghisellini et al. 1998, Maraschi et al. 2008).

\section{Discussion and summary}
The most interesting feature in the broad-band high-energy spectrum of PKS~$1830-211$ is the spectral flattening below $\sim 4$~keV. Such a flattening has been observed also by De Rosa et al. (2005) in the combined \emph{Chandra}/\emph{INTEGRAL} spectrum, but the best fit model proposed is a single power-law with $\Gamma=1.09\pm 0.05$ extending over the entire $0.5-80$~keV band absorbed by cold gas from the intervening galaxy at $z=0.89$, with column density $N_{\rm H}^z \sim 2\times 10^{22}$~cm$^{-2}$. Instead, in the present work, by analyzing a spectrum covering a wider energy range ($0.2-250$~keV), we have shown evidence of a photon deficit at low energies in addition to the absorption from the intervening galaxy, confirming and extending the results obtained with \emph{XMM-Newton} only ($0.4-10$~keV) reported by Foschini et al. (2006). This low-energy photon deficit can be best fit with a power-law harder ($\Gamma \sim 1.0$) than the one at energies greater than $\sim 4$~keV ($\Gamma \sim 1.3$).

This low-energy photon deficit has been often observed in high-redshift flat-spectrum radio quasars (Fiore et al. 1998; Worsley et al. 2004). According to Fiore et al. (1998), the photoelectric absorption intrinsic to the quasar is likely to be the origin of these low-energy roll-off. However, in the case of PKS $1830-211$, the tests performed to fit with a ionized absorber the low-energy data gave worst results (see Sect.~2.2). The broken power-law is statistically required and this suggests that the spectral break is likely to be due to the intrinsic curvature of the spectrum near the low-energy end of the External-Compton (EC) component, while the relative importance of the Synchrotron Self-Compton (SSC) component is likely to decrease due to the increasing importance of the external (broad-line region, accretion disk, other) radiation field (Tavecchio et al. 2007; Ghisellini et al. 2007). 

However, in the case of PKS~$1830-211$, the gravitational lensing should have an impact on the spectral and variability properties of the source, but it is not clear how to weight it at high-energy. These effects in the $\gamma$-ray band on distant blazars have been discussed by Combi \& Romero (1998) and Torres et al. (2002, 2003). The observed hard X-rays and, probably the soft X-rays as well, are the combination of the contributions from SSC and EC, which in turn are generated from different places. Therefore, the lensing can act differently, resulting in changes in spectral shape.

We can also probe further on the possible intrinsic spectrum and the amplification factor corresponding to different part of the energy spectrum of PKS~$1830-211$. We assume that, apart from the absorptions in Table~\ref{tab:spec-fit} due to the Galactic column and the intervening system, the spectral break and roll-off in spectrum at low energies is caused by the difference in amplification factor. Therefore, we take as the intrinsic spectrum of PKS~$1830-211$ the power law with $\Gamma \sim 1.3$ from $\sim 4.3$ to $250$~keV. We extend this value to low energies, calculate the flux in the $0.2-4.3$~keV energy band and compare to the observed one. The ratio of the amplification factor at energies between below and above $4.3$~keV is estimated by the value around $0.8$. This value appears to be too low, as it is expected that the amplification factor increases with the energy and, so far, the available modelings of the magnification factor from radio/optical observations is in the range $2-4$ (Nair et al. 1993, Swift et al. 2001, Courbin et al. 2002).

The time variability analysis could offer another way to try estimating the impact of the gravitational lensing. PKS~$1830-211$ is known to be radio variable on time scales of months (Steppe et al. 1993). Other estimations of time scales are of $44\pm 9$ days (van Ommen et al. 1995) or $24\pm 2$ days (Lovell et al. 1998). In modelling PKS~$1830-211$ by Subrahmanyan et al. (1990), the time lag between the two main components (NE and SW) is expressed as $\sim 6(z_g/0.1)(2h)^{-1}$~days, where $z_g$ is the redshift of the lensing galaxy, $h$ is the Hubble constant in units of $100$~km~s$^{-1}$~Mpc$^{-1}$. By setting the lensing galaxy at $z=0.89$, and $h \sim 75$~km~s$^{-1}$~Mpc$^{-1}$,  we have the time lag of the order of $71$~days. Therefore, the time lag between the two core components might be in the  range $24 - 70$ days. 

The hard X-ray variability displayed in the lightcurves of \emph{INTEGRAL}/ISGRI in the $20-100$~keV band and of \emph{SWIFT}/BAT in the $15-50$~keV band is quite beyond this range. The source flux can vary by a factor of $2$ on time scales of months to year and the relatively poor statistics prevent us from establishing convincing evidences of the flux variability on shorter time scales. The observed variability might be the result of the evolutions either of the relativistic jet plasma or the inverse-Compton-scattered soft target photons. It might be that, by given a steady jet plasma, the soft target photons have the density evolving over months to year, which cause the long term variability showing up at hard X-rays.  De Rosa et al. (2005) reproduced the SED of PKS 1830-211 with a SSC+EC model, where the EC component was dominated by photon field from the torus. This scenario could fit with the observed hard-X ray variability on a year time scale, when this variability is attributed to  the evolution of the soft photons target density. However, we point out that the SED built by De Rosa et al. (2005) has been corrected at high-energy by the amplification factor due to the gravitational lensing, which in turn is affected by large uncertainties, as we have already underlined. One of the main result of the amplification is to change the luminosity of the seed photon source and, therefore, the conclusions by De Rosa et al. (2005) could be severely biased by the not fully justified assumption of the amplification factor correction. A possible measurement of a time lag at hard X-rays in the future may help to resolve the contribution, if any, of the core region to the jet dominated emission.  

An indication might be that, as already pointed out in De Rosa et al. (2005), PKS~$1830-211$ is actually another blazar to have persistent MeV emissions, which are always detectable by COMPTEL.  The other two COMPTEL blazars which own MeV emission be visible over years are 3C~$273$ (Collmar et al. 2000) and 3C~$354.3$ (Zhang et al. 2005). Such a long-term steady MeV emission has been discussed in Zhang et al. (2005) for 3C~$454.3$, in the framework of leptonic multicomponent models where the MeV emission might be dominated by EC of seed photons coming directly from the accretion disk. The bulk Lorentz factor is argued to keep at a relatively high level to maintain the MeV emission visible over the time scale of years. In the case of PKS~$1830-211$, such a bulk Lorentz factor has been estimated to be about $17$ by Foschini et al. (2006), where -- given the uncertainties of the lensing at high-energies -- the modelling of the SED was performed on the observed data, without any correction.

In summary, we presented here the most updated broad-band high-energy spectrum of PKS~$1830-211$. The source presents a low-energy roll-off that can be explained efficiently in term of natural interplay between SSC and EC, as shown in other high-z FSRQ. However, it is not clear what is the weight of the amplification factor due to the gravitational lensing. Future observations at X-rays with higher spatial resolution should allow us to measure this factor.

\acknowledgments  We thank the anonymous referee for the constructive comments that were of great help in improving our paper. This work was subsidized by the National Natural Science Foundation of China, and the CAS key Project KJCX2-YW-T03. DFT acknowledges support by Spanish MEC grant AYA 2006-00530 and CSIC grant PIE 200750I029. J.-M. W. thanks the Natural Science Foundation of China for support via NSFC-10325313, 10521001 and 10733010. LF acknowledges support by ASI/INAF contract I/088/06/0.

\begin{table*}
\caption{\emph{INTEGRAL}/ISGRI observation log of  PKS~$1830-211$. The revolution ID, time in MJD, number of science windows, and the total exposure are given.}
\begin{flushleft}
\begin{tabular}{cccc}
\hline 
\multicolumn{1}{c}{Revolution ID}&\multicolumn{1}{c}{MJD }&\multicolumn{1}{c}{Number of SCW}&\multicolumn{1}{c}{Exposure (ks)}\\  \hline 
050-065 & 52710-52767 & 138  &250.9\\  
105-122 & 52875-52927 & 247  &585.1\\ 
164-186 & 53052-53119 & 108  & 205.9\\ 
225-249 & 53234-53305 & 161  &388.3\\  
286-310 & 53417-53488 & 97  &192.1\\ 
348-371 & 53602-53672 & 209  &609.1\\ 
407-432 & 53777-53854 & 135  &266.1\\
\hline
\end{tabular}\end{flushleft}
\label{tab:isgri-obs}
\end{table*}

\begin{table*}
\caption{\emph{XMM-Newton} observation log of PKS~$1830-211$. The observational ID, time in calendar date and MJD, prime observational target, the pointing offset angle and total exposure are given.}
\begin{flushleft}
\begin{tabular}{cccccc}
\hline 
\multicolumn{1}{c}{ID}&\multicolumn{1}{c}{Date}&\multicolumn{1}{c}{MJD }&\multicolumn{1}{c}{Object}&\multicolumn{1}{c}{Offset angle}&\multicolumn{1}{c}{Exposure$^{*}$}\\ 
\multicolumn{1}{c}{}&\multicolumn{1}{c}{(dd/mm/yy)}&\multicolumn{1}{c}{}&\multicolumn{1}{c}{}&\multicolumn{1}{c}{arcsecond}&\multicolumn{1}{c}{ks}\\ \hline 
0204580201 & 10/03/2004 & 53074  & PKS 1830-211 & 1.1& 3,7,7\\  
0204580301 & 24/03/2004 & 53088  & PKS 1830-211 & 1.1& 27,31,31\\ 
0204580401 & 05/04/2004 & 53100  & PKS 1830-211 & 1.1& 13,19,19\\ 
\hline
\end{tabular}
\begin{list}{}{}
\item[$^{*}$] Effective exposure on PN, MOS1, and MOS2, respectively, after having removed periods with high background. 
\end{list}
\end{flushleft}
\label{tab:xmm-obs}
\end{table*}

\begin{table*}[ptbptbptbptbptbptbptbptbptb]
\caption{Summary of the results of the individual fits on \emph{XMM-Newton} data. In all the three models we fixed the Galactic column to $N_{\rm H}=2.05\times 10^{21}$~cm$^{-2}$ and we left free an absorption redshifted to the value of the intervening galaxy ($z=0.89$); the value is in [$10^{22}$~cm$^{-2}$]. The continuum of the blazar is modeled with a redshifted power-law or with a broken power-law or with a log-parabola, according to the formula $F(E)=E^{[-a - b \log(E)]}$, where $a$ is the initial photon index and $b$ is the curvature parameter. The break energy is in [keV]. $N$ is the flux density given at $1$~keV in units of [$10^{-3}$~ph~cm$^{-2}$~s$^{-1}$~keV$^{-1}$]; $F$ is the observed flux in the $0.2-10$~keV band in units [$10^{-11}$~erg~cm$^{-2}$~s$^{-1}$].}
\vspace{1pt}
\begin{tabular}{lccccccc}
\hline
\multicolumn{8}{c}{Redshifted Power-Law}\\
ObsID        & $N_{\rm H}^{z}$ & $\Gamma$ & {} & {} & $N$ & $\tilde{\chi}^2$/dof & $F$  \\
\hline
$0204580201$ & $2.5\pm 0.2$    & $1.14\pm 0.04$ & {} & {} & $5.9\pm 0.7$ & $1.06/427$ & $1.48$\\
$0204580301$ & $2.5\pm 0.1$    & $1.12\pm 0.02$ & {} & {} & $5.2\pm 0.3$ & $1.08/1340$ & $1.38$\\
$0204580401$ & $2.2\pm 0.1$    & $1.12_{-0.02}^{+0.03}$ & {} & {} & $4.7_{-0.3}^{+0.4}$ & $1.02/867$ & $1.27$\\
\hline
\multicolumn{8}{c}{Broken Power-Law}\\
ObsID        & $N_{\rm H}^{z}$     & $\Gamma_1$ & $\Gamma_2$ & $E_{\rm break}$ & $N$ & $\tilde{\chi}^2$/dof & $F$  \\
$0204580201$ & $2.0_{-0.3}^{+0.1}$ & $0.95_{-0.13}^{+0.09}$ & $1.46\pm 0.17$ & $4.1_{-0.7}^{+0.6}$ & $1.2\pm 0.1$ & $0.99/425$ & $1.43$\\
$0204580301$ & $1.9_{-0.1}^{+0.2}$ & $0.88\pm 0.07$ & $1.26_{-0.04}^{+0.06}$ & $3.3_{-0.3}^{+0.4}$ & $1.01_{-0.06}^{+0.07}$ & $1.03/1338$ & $1.36$\\
$0204580401$ & $1.98_{-0.07}^{+0.16}$ & $1.00_{-0.03}^{+0.06}$ & $1.33_{-0.06}^{+0.19}$ & $4.3_{-0.2}^{+0.8}$ & $1.03_{-0.03}^{+0.07}$ & $0.99/865$ & $1.24$\\
\hline
\multicolumn{8}{c}{Broken Power-Law (average)$^*$}\\
{}           & $2.10 \pm 0.04$  & $1.000_{-0.018}^{+0.001}$ & $1.33_{-0.04}^{+0.03}$ & $4.26_{-0.12}^{+0.21}$ & $1.121_{-0.035}^{+0.004}$ & $1.04/1397$ & $1.34$\\
\hline
\multicolumn{8}{c}{Log-parabola}\\
ObsID & $N_{\rm H}^{z}$ & $a$ & $b$ & {} & $N$ & $\tilde{\chi}^2$/dof & $F$  \\
$0204580201$ & $1.2\pm 0.4$ & $<0.5$ & $0.77\pm 0.21$ & {} & $0.80_{-0.12}^{+0.15}$ & $0.98/426$ & $1.41$\\
$0204580301$ & $1.5\pm 0.2$ & $0.43\pm 0.12$ & $0.57\pm 0.10$ & {} & $0.81\pm 0.06$ & $1.02/1339$ & $1.34$\\
$0204580401$ & $1.5_{-0.2}^{+0.3}$ & $0.58_{-0.11}^{+0.21}$ & $0.45_{-0.09}^{+0.17}$ & {} & $0.83_{-0.06}^{+0.12}$ & $0.99/866$ & $1.26$\\
\hline
\end{tabular}
\label{tab:fitxmm}
\begin{list}{}{}
\item[$^{*}$] The average spectrum has been rebinned in order to have at least $50$ counts per bin. The PN has been considered as reference detector and the intercalibration constants are: MOS1/PN$=1.21\pm 0.01$ and MOS2/PN$=1.23\pm 0.01$.
\end{list}
\end{table*}

\begin{table*}
\fontsize{9pt}{10pt}\selectfont
\caption{COMPTEL observations of the sky region of PKS~$1830-211$ during $1991-1997$. The CGRO VPs, their time periods in calendar date and MJD, prime observational targets of \emph{CGRO}, and the pointing offset angle are given.}
\begin{flushleft}
\begin{tabular}{cccccc}
\hline 
\multicolumn{1}{c}{VP}&\multicolumn{1}{c}{Date}&\multicolumn{1}{c}{MJD }&\multicolumn{1}{c}{Object}&\multicolumn{1}{c}{Offset angle}&\multicolumn{1}{c}{CGRO Phase }\\ 
\multicolumn{1}{c}{}&\multicolumn{1}{c}{(dd/mm/yy)}&\multicolumn{1}{c}{}&\multicolumn{1}{c}{}&\multicolumn{1}{c}{}&\multicolumn{1}{c}{}\\ \hline 
5.0 & 12/07/91-26/07/91 &48449-48463& Gal. Center & 12$^{\circ}$& Phase I\\  
7.5 & 15/08/91-22/08/91 &48483-48490& Gal 025-14 & 15$^{\circ}$&\\ 
13.0 & 31/10/91-07/11/91 &48560-48567&GAL 025-14& 15$^{\circ}$&\\ 
16.0 & 12/12/91-27/12/91 &48602-48617&SCO X-1  & 29$^{\circ}$&\\ 
20.0 & 06/02/92-20/02/92 &48658-48672& SS 433   & 28$^{\circ}$& \\ 
27.0 & 28/04/92-07/05/92 &48740-48749& 4U 1543-47 & 41$^{\circ}$& \\ 
35.0 & 06/08/92-11/08/92 &48840-48845& ESO 141-55 & 41$^{\circ}$&\\ 
38.0 & 27/08/92-01/09/92 &48861-48866& ESO 141-55 & 41$^{\circ}$& \\  
42.0 & 15/10/92-29/10/92 &48910-48924& PKS 2155-304 & 40$^{\circ}$& \\ 
43.0 & 29/10/92-03/11/92 &48924-48929&MRK 509   & 29$^{\circ}$& \\ \hline
209.0 & 09/02/93-22/02/93 &49027-49040&2CG010-31 & 30$^{\circ}$& Phase II\\  
210.0 & 22/02/93-25/02/93 &49040-49043& Gal. Center & 20$^{\circ}$&\\ 
214.0& 29/03/93-01/04/93 &49075-49078&Gal. Center& 20$^{\circ}$&\\ 
219.4 & 05/05/93-06/05/93 &49112-49113&Gal. Center  & 31$^{\circ}$&\\ 
223.0 & 31/05/93-03/06/93 &49138-49141&Gal. Center   & 14$^{\circ}$& \\ 
226.0 & 19/06/93-29/06/93 &49157-49167&Gal. 355+5  & 20$^{\circ}$& \\ 
231.0& 03/08/93-10/08/93 &49202-49209&  NGC 6814  & 12$^{\circ}$&\\ 
229.0& 10/08/93-11/08/93 &49209-49210&Gal. 5+05  & 13$^{\circ}$& \\  
229.5 & 12/08/93-17/08/93 &49211-49216&Gal. 5+05 & 13$^{\circ}$& \\ 
232.0& 24/08/93-26/08/93 &49223-49225&Gal. 348-00   & 25$^{\circ}$& \\ 
232.5& 26/08/93-07/09/93 &49225-49237&Gal. 348-00  & 25$^{\circ}$& \\ \hline
302.3 & 09/09/93-21/09/93 &49239-49251&GX 1+4  & 18$^{\circ}$& Phase III\\  
323.0 & 22/03/94-05/04/94 &49433-49447&Gal. 357-112 & 16$^{\circ}$&\\ 
324.0& 19/04/94-26/04/94 &49461-49468&Gal. 016+05 & 12$^{\circ}$&\\ 
330.0 & 10/06/94-26/04/94 &49513-49517&Gal. 018+00    & 8$^{\circ}$&\\ 
332.0 & 18/06/94-05/07/94 &49521-49538&Gal. 018+00  & 8$^{\circ}$& \\ 
334.0 & 18/07/94-25/07/94 &49551-49558&Gal. 009-08  & 4$^{\circ}$& \\ 
336.5& 04/08/94-09/08/94 & 49568-49573&GRO J1655-40& 33$^{\circ}$&\\ 
338.0& 29/08/94-31/08/94 &49593-49595&GRO J1655-40& 28$^{\circ}$& \\  
339.0& 20/09/94-04/10/94 &49615-49629& 3C 317 & 47$^{\circ}$& \\\hline 
414.3& 29/03/95-04/04/95 &49805-49811&GRO J1655-40   & 26$^{\circ}$& Phase IV \\ 
421.0& 06/06/95-13/06/95 &49874-49881&Gal. Center   & 18$^{\circ}$& \\ 
422.0 & 13/06/95-20/06/95 &49881-49888& Gal. Center & 18$^{\circ}$&\\  
423.0 & 20/06/95-30/06/95 &49888-49898&Gal. Center & 11$^{\circ}$&\\ 
423.5& 30/06/95-10/07/95 &49898-49908&PKS 1622-297& 32$^{\circ}$&\\ 
429.0 & 20/09/95-27/09/95 &49980-49987&GAL 018+04   & 11$^{\circ}$&\\ \hline
501.0 & 03/10/95-17/10/95 &49993-50007&GAL 28+4 & 18$^{\circ}$& Phase V\\  
508.0 & 14/12/95-20/12/95 &50065-50071&GAL 5+0 & 9$^{\circ}$&\\ 
509.0& 20/12/95-02/01/96 &50071-50084&GAL 21+14& 21$^{\circ}$&\\ 
513.0 & 06/02/96-13/02/96 &50119-50126&PKS 2155-304  & 47$^{\circ}$&\\ 
516.1 & 18/03/96-21/03/96 &50160-50163&GRO J1655-40  & 33$^{\circ}$& \\ 
520.4 & 21/05/96-28/05/96 &50224-50231&PKS 2155-304 & 47$^{\circ}$& \\ 
524.0& 09/07/96-23/07/96 &50273-50287&GX 339-4 & 29$^{\circ}$&\\ 
529.5& 27/08/96-06/09/96 &50322-50332&GRO J1655-40 & 28$^{\circ}$& \\ \hline 
624.1& 04/02/97-11/02/97 &50483-50490&Gal 16+00   & 10$^{\circ}$&Phase VI \\ 
619.2& 14/05/97-20/05/97 &50582-50588&GRS 1915+105    & 35$^{\circ}$& \\ 
620.0 & 10/06/97-17/06/97 &50609-50616&Gal 16+04 & 10$^{\circ}$& \\  
625.0 & 05/08/97-19/08/97 &50665-50679&GRS 1758-258  & 13$^{\circ}$&\\ 
615.1& 19/08/97-26/08/97 &50679-50686&PKS 1622-297& 30$^{\circ}$&\\ \hline

\end{tabular}\end{flushleft}
\label{tab:mev-obs}
\end{table*}

\begin{table}[ptbptbptb]
\caption{Fluxes of PKS~$1830-211$ and the time periods of COMPTEL observations are listed. The error bars are $1\sigma$.}
\begin{flushleft}
\begin{tabular}{cccccc}\hline
\multicolumn{1}{c}{Period}&\multicolumn{1}{c}{MJD}&\multicolumn{4}{c}{Flux (10$^{-5}$ ph cm$^{-2}$ s$^{-1}$)}\\  
\multicolumn{1}{c}{ }&\multicolumn{1}{c}{} &\multicolumn{1}{c}{0.75-1 MeV}&\multicolumn{1}{c}{1-3  MeV}&\multicolumn{1}{c}{3-10 MeV}&\multicolumn{1}{c}{10-30 MeV}\\  \hline
Phase 1-6  &48392-50539  &4.9$\pm$1.4   &7.6$\pm$1.4   &1.5$\pm$0.6   &0.4$\pm$0.3 \\  \hline
\end{tabular}\end{flushleft}
\label{tab:mev-flux}
\end{table}

\begin{table}[ptbptbptbptbptbptbptbptbptb]
\caption{Results of the joint \emph{XMM-Newton}/\emph{INTEGRAL} data ($0.2-250$~keV) fit with a broken power-law model plus two absorption components. Absorption columns are in units [$10^{22}$~cm$^{-2}$]; the break energy is in [keV]; the normalization $N$ is in [$10^{-3}$~ph~cm$^{-2}$~s$^{-1}$~keV$^{-1}$]. The intercalibration constant between EPIC-PN and ISGRI is ISGRI/PN = 0.63$\pm$0.07.}
\vspace{1pt}
\label{10_spe}
\begin{tabular}{c|cc|cccc|c}
\hline
\multicolumn{1}{c|}{wabs}
&\multicolumn{2}{c|}{zwabs}
&\multicolumn{4}{c|}{bknpow}
& \\
$N_{H}$ & $N_{H}$    & $z$   &$\Gamma_{1}$    & $E_{\rm break}$   & $\Gamma_{2}$     & $N$  & $\tilde{\chi}^2$/dof \\
\hline
$0.205$ (fix) & $1.96\pm0.09$ & $0.886$ (fix)  & $0.93^{+0.04}_{-0.03}$  &$3.68^{+0.29}_{-0.19}$ & $1.29\pm0.04$ & $1.05\pm0.04$ & $1.04/1406$ \\ 
\hline
\end{tabular}
\label{tab:spec-fit}
\end{table}

\end{document}